\documentstyle[twoside,fleqn,espcrc2,epsfig]{article}
\def\lesssim{\lower.7ex\hbox{${\buildrel < \over \sim}$}}
\def\gtrsim{\lower.7ex\hbox{${\buildrel > \over \sim}$}}

\newcommand{\AmS}{{\protect\the\textfont2
  A\kern-.1667em\lower.5ex\hbox{M}\kern-.125emS}}
\title{Atmospheric Neutrino Fluxes}

\author{Thomas K. Gaisser\address{Bartol Research Institute,
        University of Delaware,\\
        Newark, DE 19716, USA}%
        \thanks{Research supported in part by the U.S. Department
            of Energy under Grant No. DE-FG02-91ER40626}}

\begin{document}
\begin{abstract}
This talk is a status report on calculations of the flux of 
atmospheric neutrinos from the sub-GeV range
to $E_\nu\sim$~PeV.   In the lower energy range ($E_{\nu}< 1$~TeV)
the primary interest is in using the atmospheric neutrino beam
to study neutrino oscillations.  In the TeV range and above,
atmospheric neutrinos are a calibration source and background
for neutrino telescopes.

\end{abstract}

\maketitle

\noindent
\section{INTRODUCTION}

The discovery of neutrino oscillations with atmospheric neutrinos
makes it important to know the production spectrum of neutrinos
as precisely as possible in order to infer the properties and
parameters of the oscillations from the data.  It also means
that the flux of cosmic-ray induced neutrinos is much better measured
than it otherwise might have been.  In addition to the extensive
measurements at Super-K,~\cite{SuperK} there were important measurements
at Soudan~\cite{Soudan}
and MACRO.~\cite{MACRO} 
The measurements cover a range of energies and techniques,
and they see the beam from different locations in the geomagnetic
field, which exposes interesting effects at low energy.

Because of oscillations, measuring the cosmic-ray neutrino beam
is an iterative process in which the oscillations and fluxes
must be understood from the same data.  Fortunately, calculation
of the neutrino spectrum at production is straightforward, and
it can be checked by comparison to measurements of atmospheric muons.
Moreover, the evidence for oscillations
is robust because it is based on ratios, which are better
known that the absolute normalization of the atmospheric neutrino beam.
The anomalous ratio
of electron-like to muon like events reveals a relative 
deficit of $\nu_\mu$ at low energy, and
the ratio of upward to downward multi-GeV events reflects the
pathlength dependence of $\nu_\mu$ oscillations and defines a
range of $\delta m^2$.~\cite{SuperK}   
All this is reinforced by the low ratio
of stopping to throughgoing upward, neutrino-induced 
muons~\cite{M2,SK2} and by the
low ratio of vertically upward to horizontal 
throughgoing muons.~\cite{M3,SK3}
A consistent pattern of energy and pathlength dependence
emerges that clearly points to neutrino oscillations as the explanation.
Because atmospheric $\nu_e$ behave normally to the precision
measured so far, the main effect must lie in the
$\nu_\mu\leftrightarrow\nu_\tau$ sector (or involve sterile neutrinos, which 
are now disfavored).~\cite{SK4,MACRO}

Here I review the main features of the calculation of the atmospheric
neutrino beam at production, emphasizing the simple features that provide 
the basis for the evidence for oscillations.  This talk is based to
a large extent on a recent review~\cite{GH}.  I organize the material
here in order of increasing energy.

\noindent
\begin{figure}[t]
\vspace{-.5cm}
\flushleft{\epsfig{figure=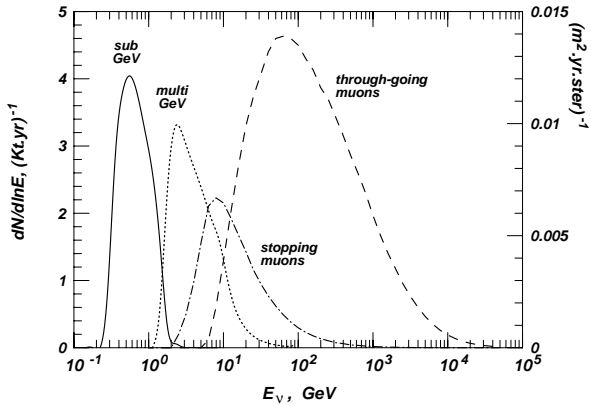,width=8.5cm}}
\label{fig1}
\vspace{-1cm}
\caption{Distribution of neutrino energies that give rise to several
classes of neutrino events.}
\end{figure}

\noindent
\begin{figure}[h]
\flushleft{\epsfig{figure=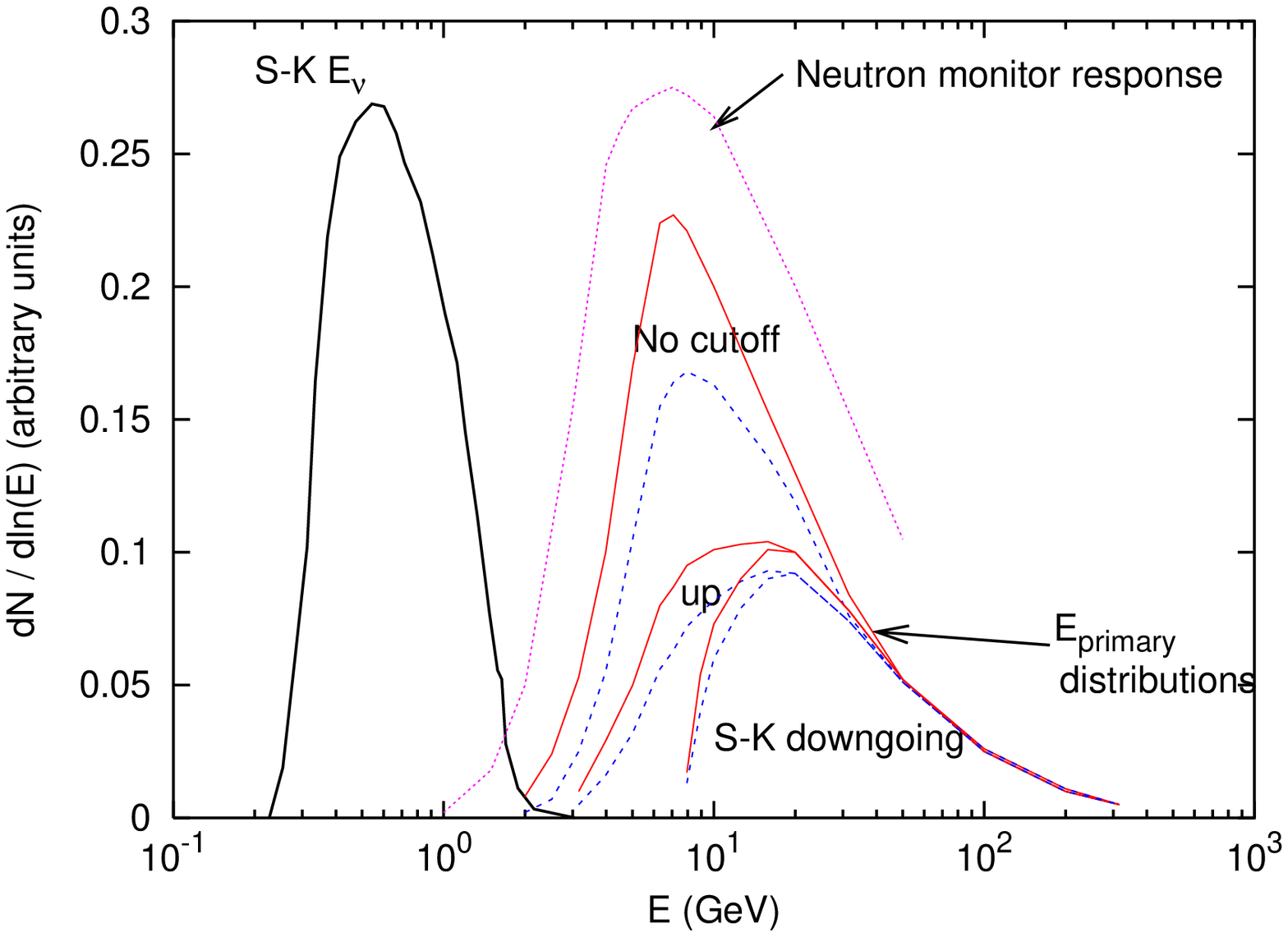,width=7.5cm}}
\label{fig2}
\vspace{-1cm}
\caption{Response functions for sub-GeV neutrinos under several conditions
(see text).  The three pairs of curves show response functions
for the distribution of neutrino energies in the solid 
curve (S-K $E_\nu$), which is the same as the left-most curve of
 Fig.~1.}
\end{figure}

The atmospheric neutrino flux is a convolution of the primary spectrum at
the top of the atmosphere with the yield ($Y$)
of neutrinos per primary particle.  
To reach the atmosphere and interact, the primary cosmic rays first
have to pass through the geomagnetic field.
Thus the flux of neutrinos of type $i$ can be represented as
\begin{eqnarray}
\label{nuflux}
\phi_{\nu_i} & = & \phi_p\,\otimes\,R_p\,\otimes\,Y_{p\rightarrow\nu_i}\\
\nonumber
 &  & +\;\sum_A \left\{\phi_A\,\otimes\,R_A\,\otimes\,Y_{A\rightarrow\nu_i}
\right\},
\end{eqnarray}
where $\phi_{p(A)}$ is the flux of primary protons (nuclei of mass A)
outside the influence of the geomagnetic field
and $R_{p(A)}$ represents the filtering effect of the geomagnetic field.
Free and bound nucleons are treated separately.

Each of the factors on the right side of Eq.~\ref{nuflux} is a potential
source of uncertainty.  Since the uncertainties depend on energy, one needs an
estimate of the relative importance of different primary energies for
a given region of neutrino energy.  Fig.~1~\cite{Engel}
shows the distributions of neutrino energies for four classes of events.
Very roughly, sub-GeV events, multi-GeV events, upward stopping muons
and upward throughgoing muons correspond respectively to
primary cosmic-ray 
energies of $10^{1\pm0.5}$, $10^{1.5\pm0.5}$, $10^{2.0\pm0.5}$ 
and $10^{3.0\pm1}$~GeV.  Fig.~2 shows the response function
in detail for the sub-GeV events, as defined at Super-K~\cite{SuperK}.

\noindent
\begin{figure}[t]
%\vspace*{-1cm}
\flushleft{\epsfig{figure=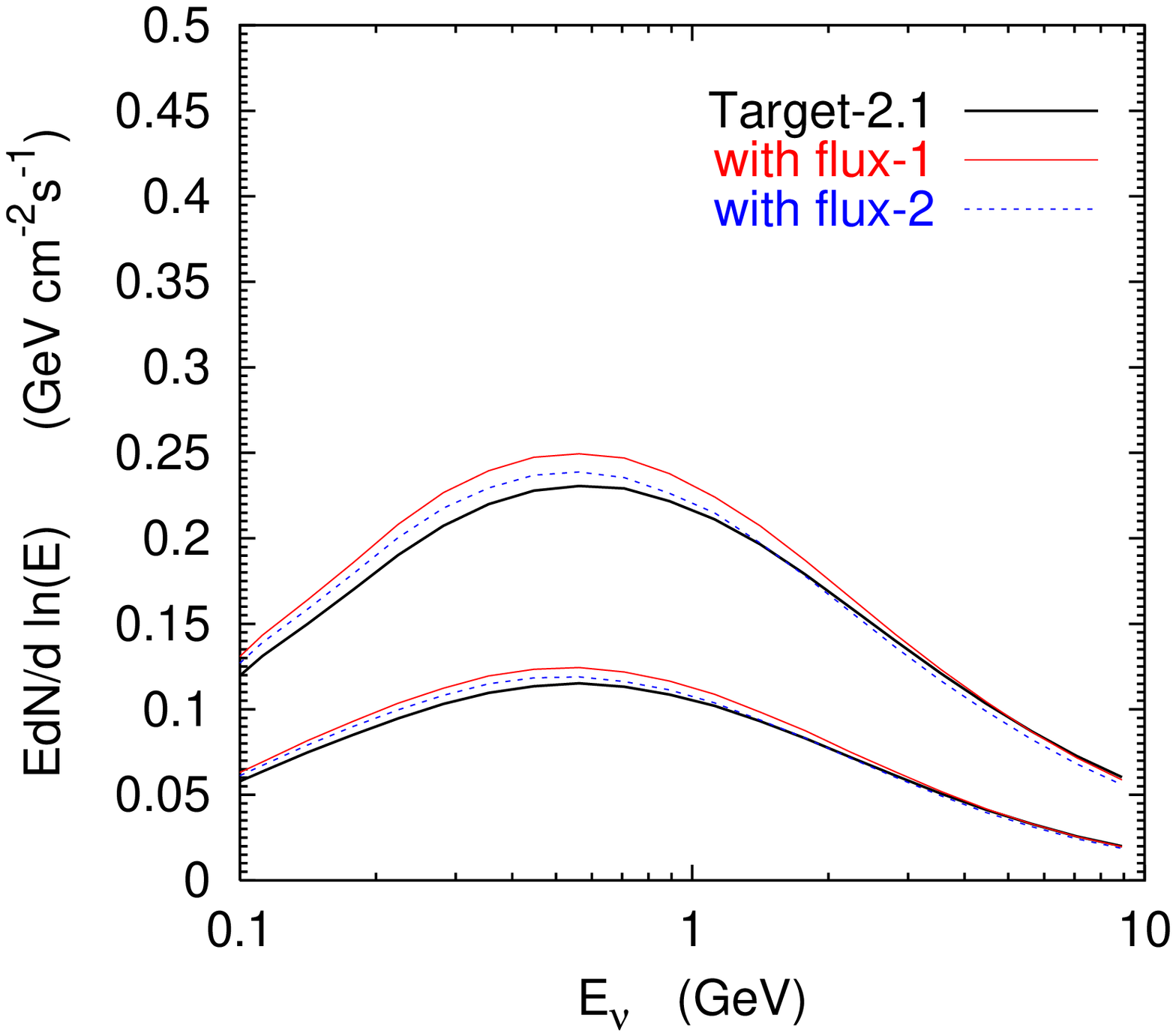,width=7.5cm}}
\label{fig3}
\vspace{-1cm}
\caption{Comparison of neutrino flux for three primary spectra
for the location of Kamioka.}
\end{figure}

\noindent
\begin{figure}[t]
%\vspace*{-1cm}
\flushleft{\epsfig{figure=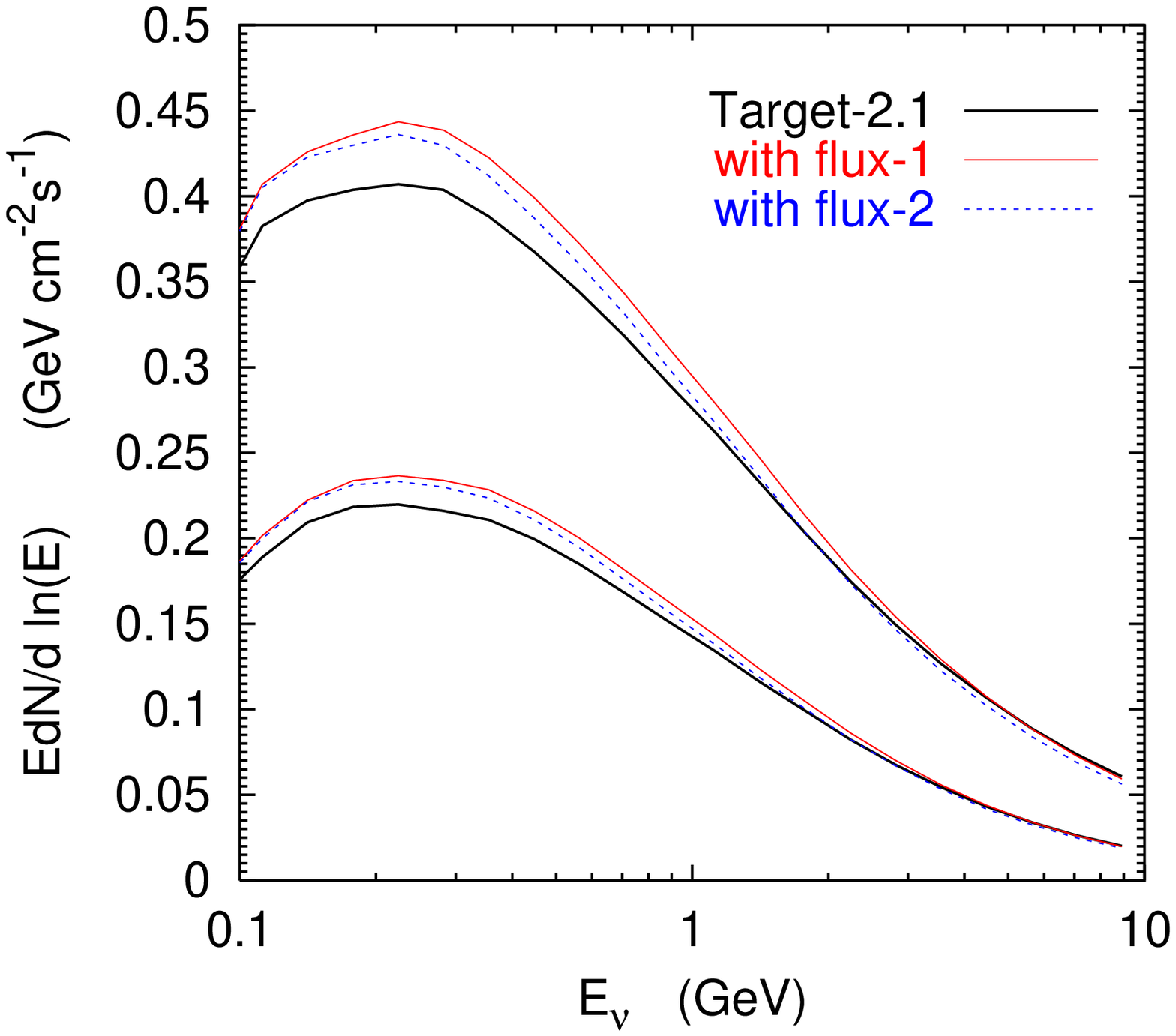,width=7.5cm}}
\label{fig4}
\vspace{-1cm}
\caption{Comparison of neutrino flux for three primary spectra
for the location of Soudan.}
\end{figure}

\section{LOW ENERGIES}

The sub-GeV
response is at low enough energy
so that it depends strongly on geomagnetic location and to some extent on the
state of solar modulation, as illustrated in Fig.~2.
  The three pairs of
curves show the 
distributions of primary energy per nucleon that produce
sub-GeV events under different conditions.  These response
curves are normalized so that the area under each curve
is proportional to the corresponding signal in the absence 
of oscillations.
``S-K downgoing" is for events at Super-K induced by cosmic-rays from above;
``up" is for events at Super-K induced by cosmic-rays 
from below the horizon; and
``No-cutoff" is the response in the absence of any
geomagnetic field effects.  The latter situation is realized
in practice for downward events at Soudan and SNO, 
which have low local geomagnetic
cutoffs.  The local geomagnetic cutoff at Super-K is
high (11.5 GV for vertically downward events), so the rate of
downward sub-GeV events at Super-K in the absence of oscillations
would be significantly lower than the rate of upward events.
This effect is partially washed out by the angle between the
direction of the charged lepton, which is measured, and the 
neutrino that produced it, and it is reversed for muon-like
events by the effect of neutrino oscillations~\cite{SuperK}.

Solar modulation may also have a noticeable effect on rates in
a sufficiently large statistical sample.  The dashed lines
show the maximum effect of solar modulation (corresponding to
conditions during the last two years of data taking at Super-K).
Neutron monitor records give an appropriate measure of
solar modulaton because the neutron monitor response is
similar to the response of atmospheric neutrinos.  The shape
of the response function of a high-latitude neutron monitor
is indicated in Fig.~2.~\cite{Clem}

\subsection{East-West effect}

This effect arises from the systematic bending
of the positively charged primary cosmic rays in the geomagnetic
field before they reach the atmosphere to interact.   At low
geomagnetic latitudes the result
is a significantly larger geomagnetic cutoff for particles arriving
from the East than for those arriving from the West.  
Thus the integrated cosmic-ray intensity is greater from
the West than from the East, and this asymmetry
is reflected by secondary neutrinos and muons.  Observation
of the East-West effect in muons confirmed long ago that the primary
cosmic rays are positive.~\cite{Johnson,Alvarez}  
Now observation of this effect with neutrinos at  
Super-Kamiokande~\cite{SuperK-EW} provides an
important systematic confirmation of the analysis and interpretation
of the evidence for neutrino oscillations.  

Single-ring events with momentum in the interval 0.4 to 3 GeV/c
and zenith angles ($\theta$) within $30^\circ$ of the horizon were 
compared~\cite{SuperK-EW} to calculations~\cite{LSG,HKKM}
of the shape of the distribution in azimuth ($\phi$).
The comparison
is independent of oscillation effects to first order because
the distribution of pathlengths is constant over the $360^\circ$
band of $\phi$ with $60^\circ < \theta < 120^\circ$.
Thus the agreement between observed and expected shapes
shows that geomagnetic effects are under control in the
atmospheric neutrino flux calculations, reinforcing confidence
in the interpretation of the up-down asymmetry of the flux
of muon neutrinos as a consequence of oscillations.

\subsection{Primary spectrum}
An estimate~\cite{spectrum} 
of the uncertainty in the primary spectrum in the
light of recent measurements by BESS~\cite{BESS} 
and AMS~\cite{AMS} is
$\pm5$\% below $100$~GeV/nucleon, which covers
the energies relevant for sub-GeV events.
This estimate is
based on a fit to BESS and AMS data alone.
A more proper estimate of the uncertainty in the primary
spectrum would use all valid measurements,
including those with larger quoted uncertainties, to estimate the systematic
uncertainty in the primary spectrum.  
Excluding an early measurement~\cite{Webber} which appears
to be anomalously high, the measurements cover a range of
$\sim~+10$\% to $-20$\% below 100~GeV relative to the measurements
of BESS and AMS.

Figs.~3 and 4 illustrate the uncertainty in the neutrino
flux from the primary spectrum uncertainty.  The plots show the
calculations using three fits to the primary spectrum
with the same interaction
model and cascade calculation~\cite{AGLS}.  Fig.~5 shows
the three assumed spectra.  The fit of Agrawal {\it et al.}~\cite{AGLS}
was made before the measurements of BESS, AMS and
other recent measurements~\cite{CAPRICE,IMAX} and reflects
mainly the earlier measurement~\cite{LEAP}, 
which has a lower normalization.  The two new fits~\cite{spectrum}
are dominated by the data from BESS and AMS.
The plot shows the all-nucleon spectrum,
which includes the nucleons bound in nuclei. 

\noindent
\begin{figure}[t]
%\vspace*{-1cm}
\flushleft{\epsfig{figure=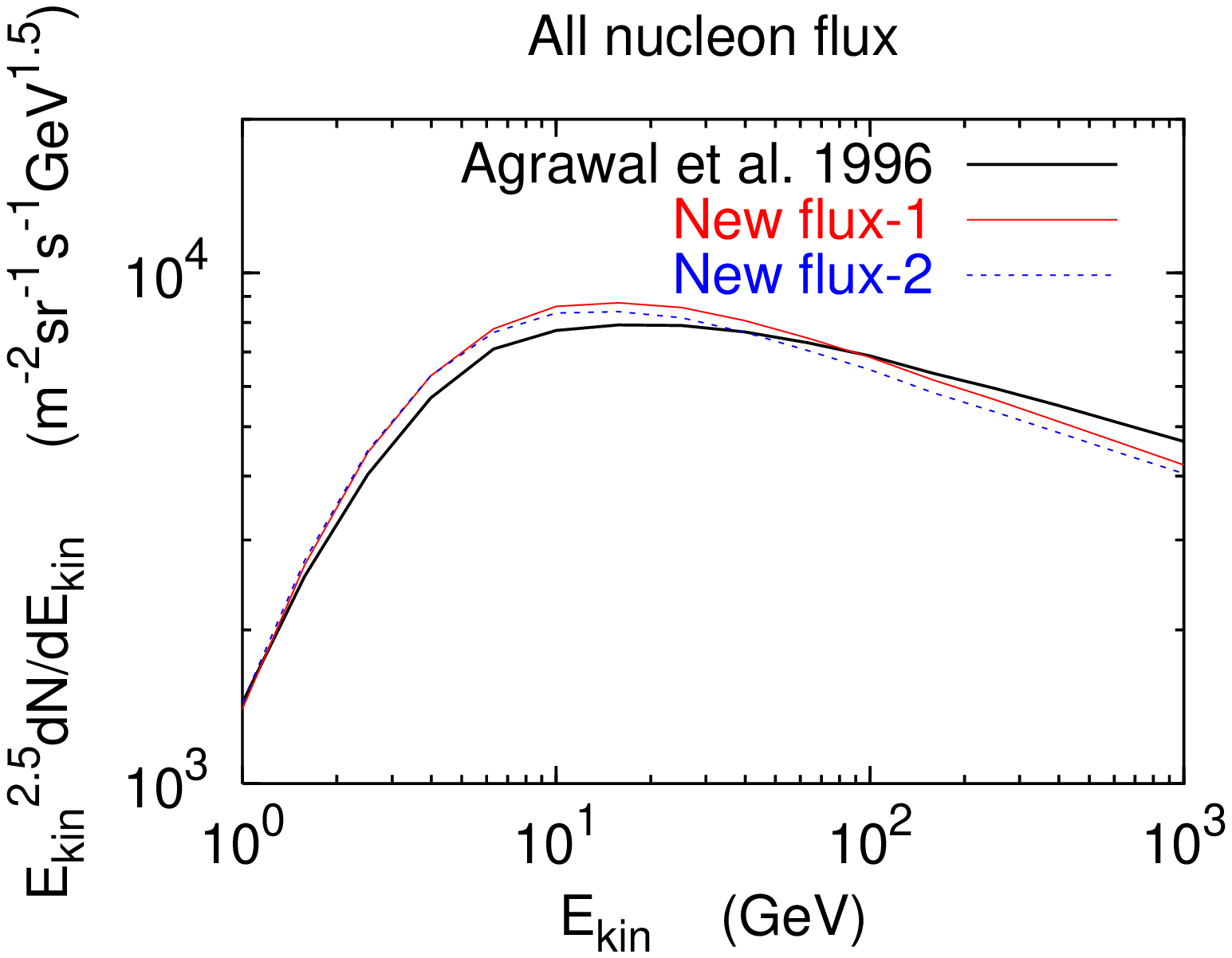,width=7.5cm}}
\label{fig5}
\vspace{-1cm}
\caption{Three fits to the spectrum of cosmic-ray nucleons.}
\end{figure}

\subsection{Treatment of hadronic interactions}
Atmospheric neutrinos come from decay of pions and kaons produced by 
interactions of cosmic-ray hadrons in the atmosphere.  
The neutrino flux depends primarily on the inclusive cross
section for $p\,+\,air\rightarrow \pi^\pm\,+\,X$.  Although pion
production is peaked near $x_{F}=0$, because of the steep primary 
spectrum the most important region of
phase space for the convolution~\ref{nuflux} 
is in the forward fragmentation region.
For pions the range $0.1<x_F<0.5$ has most weight, while for kaons the
range extends to somewhat larger $x_F$ because
of $p\rightarrow \Lambda + K^+$.  

Event generators used in calculations of atmospheric neutrinos
are based on various accelerator data.  Especially important
are several spectrometer experiments with protons in
the momentum range $15$--$30$~GeV/c incident on light nuclei.
Each experiment has statistical and systematic uncertainties.
In addition, there are systematic uncertainties associated with
extrapolation into unmeasured regions of phase space~\cite{Engel}.
Comparison of calculations made with different interaction
models using the same primary
spectrum gives an indication of the magnitude of uncertainties
that arise from treatment of hadronic interactions.  Such a comparison
is shown in Figs.~6 and 7 made with the primary spectrum
of Ref.~\cite{AGLS} (heavy solid line in Fig.~5).

Figs.~6 and 7 from Ref.~\cite{GH}
display three independent calculations
using five different representations of hadronic interactions.
The event generator Target-1 was used in the Bartol 
neutrino flux calculation~\cite{AGLS}.  Target-2.1
is a modified version, which is under development
for use in three-dimensional calculations.  
The event generator has been adjusted
to give better agreement with the pion data around $X_{lab}=0.2$.  Distributions of leading nucleons have been adjusted 
to give a better representation of neutron/proton ratios.  A preliminary description
of this work, which is still in progress, is given in Ref.~\cite{Hamburg}.
HKKM indicates the interaction model used in the earlier calculation of 
Honda {\it et al.}~\cite{HKKM}; currently~\cite{Hondanew} they use
the event generator Dpmjet3~\cite{Dpmjet}.  
FLUKA refers to the
three-dimensional calculation of Battistoni {\it et al.}~\cite{Battistoni}.
The differences are primarily due to the different representations
of pion production, rather than to technical differences.

Differences
among the calculations are significantly 
larger at Soudan than at Kamioka, which points to large
differences in the treatment of low energy interactions
($<10$~GeV), which are mostly below the geomagnetic cutoff at Kamioka.  
Forthcoming results from the HARP experiment~\cite{HARP} should be helpful
in reducing this ambiguity.  This should allow a useful, quantitative
comparison between Soudan and Kamioka, taking
account of the rather different geomagnetic environments.

\noindent
\begin{figure}[t]
%\vspace*{-1cm}
\flushleft{\epsfig{figure=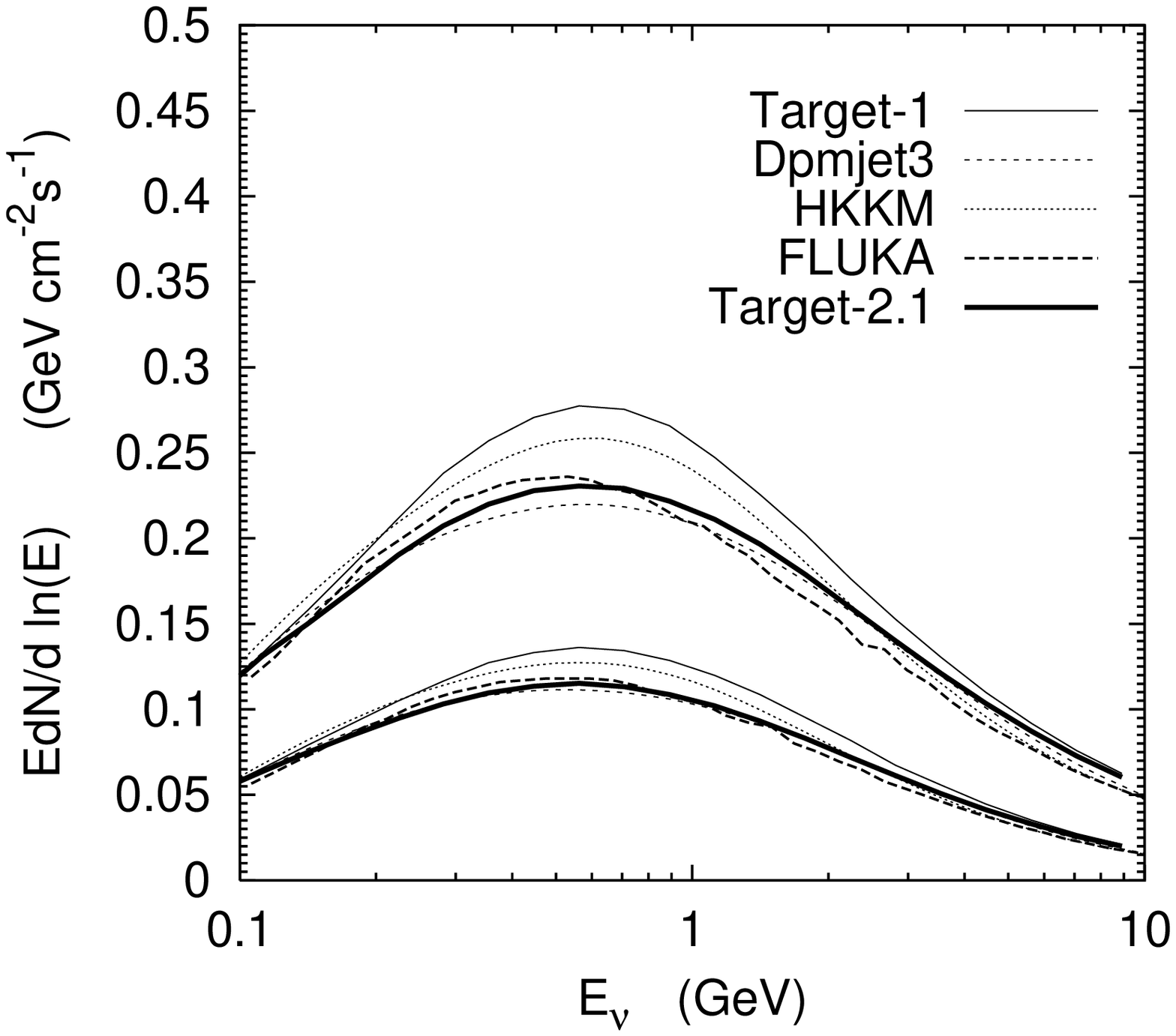,width=7.5cm}}
\label{fig6}
\vspace{-1cm}
\caption{Comparison of atmospheric neutrino flux calculations
for the location of Kamioka.}
\end{figure}

\noindent
\begin{figure}[t]
%\vspace*{-1cm}
\flushleft{\epsfig{figure=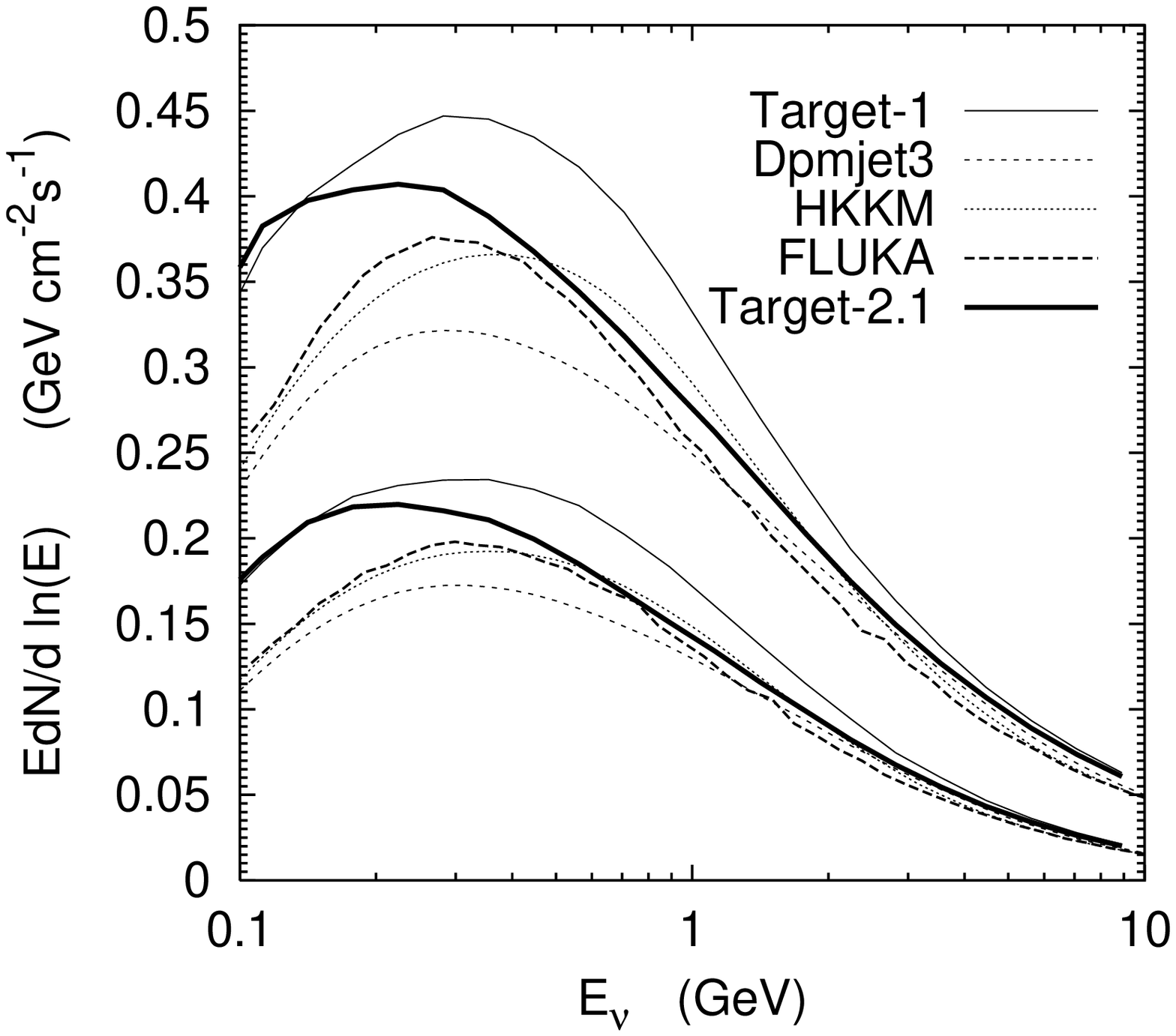,width=7.5cm}}
\label{fig7}
\vspace{-1cm}
\caption{Comparison of atmospheric neutrino flux calculations
for the location of Soudan.}
\end{figure}

\subsection{Three-dimensional calculations}

An approximation common to the calculations used to interpret
the measurements of atmospheric neutrinos so far is
that the neutrinos follow the direction of the primaries that
produce them.  This approximation becomes questionable at 
energies low enough so the typical transverse momenta of charged pions
are comparable to their longitudinal momenta, 
i.e. for $E_\nu<1$~GeV.  Another aspect of one-dimensional
calculations is that the bending of secondary charged particles
in the geomagnetic field is also neglected.  For a fully three-dimensional
calculation, one needs to generate events from an isotropic 
distribution of primaries over the full primary energy spectrum
(from pion production threshold) at a dense grid of locations
uniformly distributed over the surface of the Earth, while
accounting for the angular dependence of the geomagnetic
cutoffs at each location.  In that 
case only a tiny fraction ($\sim A/R_\oplus^2$) of the neutrinos generated
will cross a detector of area $A$.  Because the radius of the
Earth $R_\oplus\gg\sqrt{A}$, a brute force calculation is a
challenge.  

So far, two types of approximation have been used for
three-dimensional calculations.
Battistoni {\it et al.}~\cite{Battistoni} neglect bending
of secondary particles in the geomagnetic field.  In this
case, a set of cascades generated at a single location (with no
cutoff) can be used to obtain the neutrino flux from all
directions.  This is done by placing the detector where
each neutrino crosses the depth of the detector (once as it enters
and once as it exits the Earth).  The neutrino is then kept
if the direction of the primary is such that it would have
passed the geomagnetic field coming from the direction required
to make the neutrino pass through the detector at its true location.
The other approximation (e.g.~\cite{Lipari1,Honda3D,Tserk}) 
is to make $A$ large.
There are also preliminary versions of more ambitious 
calculations~\cite{Wentz,Plyaskin}.

An excess of low energy neutrinos from near the horizon is a 
characteristic feature of three-dimensional
calculations.~\cite{Battistoni}
The origin of this geometrical effect is discussed by Lipari.~\cite{Lipari1}
Because it occurs only for low energy neutrinos, however, the excess
is washed out by the broad angular distribution of the
the charged lepton relative to the neutrino direction.  

Associated with
the horizontal excess is a small change in the pathlength
distribution of low energy neutrinos.  Quantitatively, the
average pathlength of neutrinos arriving from just above the
horizon ($80^\circ < \theta < 90^\circ$)
is about 10\% lower in the 3D calculation
than in the 1D approximation for $E_\nu=0.3$~GeV.~\cite{Honda3D} 
The suppression for $E_\nu=1$~GeV is $<5$\%.  This effect, which has
not yet been included in analysis of Super-K, could be expected 
to increase the inferred $\delta m^2$ slightly (of order $\sim 1$\%
to keep the factor $\delta m^2 \times L / E_\nu$ constant while
decreasing $L$ by $\sim10$\% for $\sim10$\% of sub-GeV events).

During its test flight on the Space Shuttle, AMS mapped the
energy spectrum of low-energy primary cosmic rays around the
globe.  The data show a substantial flux of protons
with energies below the local geomagnetic cutoff.~\cite{AMS2}
These are secondary protons produced by interactions of
cosmic-rays above the cutoff that enter the atmosphere at
large zenith angles.  Depending on their orientation relative
to the geomagnetic field, a large fraction of the sub-cutoff
secondaries curve back into space.~\cite{Treiman}   Such albedo
particles may remain trapped for several cycles before they
re-enter the atmosphere.  As a consequence, the rate per
area-solid angle at which they re-enter is
correspondingly lower than their flux at 380~km. 
These features have been evaluated quantitatively by Zuccon
{\it et al.}~\cite{zuccon}.  Since both the flux of re-entering
subcutoff particles and their pion multiplicity is lower
than for primaries above the cutoff, their contribution to
the neutrino flux should be small.  It is interesting to note
that all sub-cutoff particles are included in the
1D calculations because all secondaries are assumed to follow
the direction of the incident primary.  In this respect, the 1D
calculation
overestimates slightly the contribution of sub-cutoff particles since
a few may interact on the way up rather than upon re-entry.

Another consequence of bending of secondary charged particles
in the geomagnetic field, in this case involving muons, is a
second-order systematic effect on the calculation of the East-West
effect.  The asymmetry is increased for $\nu_e$ and reduced slightly for $\nu_\mu$.~\cite{Lipari2}

\section{HIGHER ENERGY}
For $E_\nu >1$~GeV the energies of the parent cosmic rays
are sufficiently high that geomagnetic effects and solar
modulation become unimportant.  In addition, the direction
of the charged lepton is better aligned with the direction
of the neutrino.  Thus for multi-GeV
events and neutrino-induced muons, in the absence of
oscillations, one sees the characteristic enhancement 
of the neutrino flux from near the horizontal.  This is
related to the "secant theta effect" familiar from
atmospheric muons.  It is a consequence of the enhanced
decay probability for pions at large zenith angles.
The corresponding effect for neutrinos
is nicely illustrated in the plots of
angular distributions of neutrino events from Super-K.~\cite{SuperK}

\subsection{Angular dependence}

An analytic approximation for
the differential flux
of $\nu_\mu + \bar{\nu}_\mu$ from decay of pions and kaons displays
some important features the neutrino flux.
\begin{eqnarray} \nonumber
{dN_\nu\over dE_\nu}&=&{\phi_N(E_\nu)\over (1 - Z_{NN})(\gamma+1)}\left\{
\left[{Z_{N\pi}(1-r_\pi)^\gamma\over1+B_{\pi\nu}\cos\theta 
E_\nu/\epsilon_\pi}\right]\right. \\ 
&+& \left. 0.635\left[{Z_{NK}(1-r_K)^\gamma\over1+B_{K\nu}\cos\theta 
E_\nu/\epsilon_K}\right]
\right\},
\label{analytic}
\end{eqnarray}
where
\begin{equation}
\phi(E_0)\;=\;{dN\over dE_0} \;=\;A\times E_0^{-(\gamma+1)}
\label{power}
\end{equation} 
is the differential primary spectrum of nucleons of energy $E_0$.  
The neutrino flux in Eq.~\ref{analytic} is proportional to the
primary spectrum evaluated at the energy of the neutrino.
The constants $r_i=m_\mu^2/m_i^2$ for $i=(\pi,K)$, and the
constants $B_i$ depend on hadron attenuation lengths
as well as decay kinematics~\cite{book}.  

The critical energy
for pions is $\epsilon_\pi \approx 115$~GeV, while for kaons
$\epsilon_K\approx 850$~GeV.  For $E_\nu \ll \epsilon_\pi/\cos\theta$ 
the neutrino spectral index is the same as that of the primary
spectrum.  For $E_\nu>\epsilon_\pi/\cos\theta$ the spectrum steepens, first
for neutrinos from pion decay and at higher energy
($E_\nu>\epsilon_K/\cos\theta$) for neutrinos from decay of kaons.
As energy increases so does the horizontal enhancement from
the zenith angle dependence in the denominator of Eq.~\ref{analytic}.
Fig.~8 shows the fraction of atmospheric neutrinos
and atmospheric muons that come from pions and from kaons at two angles.
Above $\sim 100$~GeV, kaons dominate as the source of neutrinos.
It is the larger value of $\epsilon_K$,
together with the form of the kinematical factor in the numerator,
that determines the behavior shown in Fig.~8.

\noindent
\begin{figure}[t]
%\vspace*{-1cm}
\flushleft{\epsfig{figure=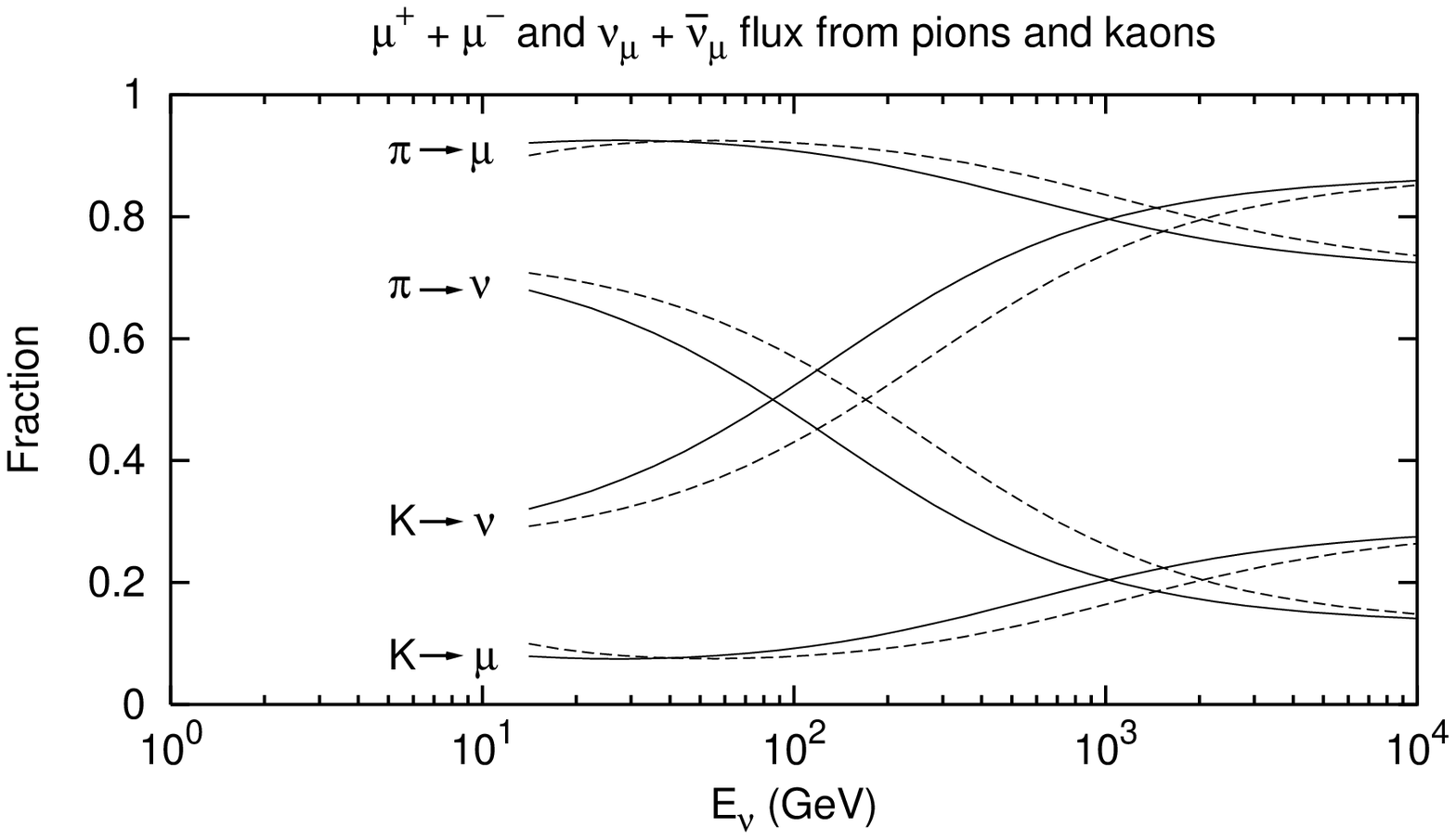,width=8.cm}}
\label{fig8}
\vspace{-1cm}
\caption{Fraction of atmospheric muons and neutrinos from
pions and kaons.  Solid: vertical; dashed: $60^\circ$.}
\end{figure}

\noindent
\begin{figure}[t]
%\vspace*{-1cm}
\flushleft{\epsfig{figure=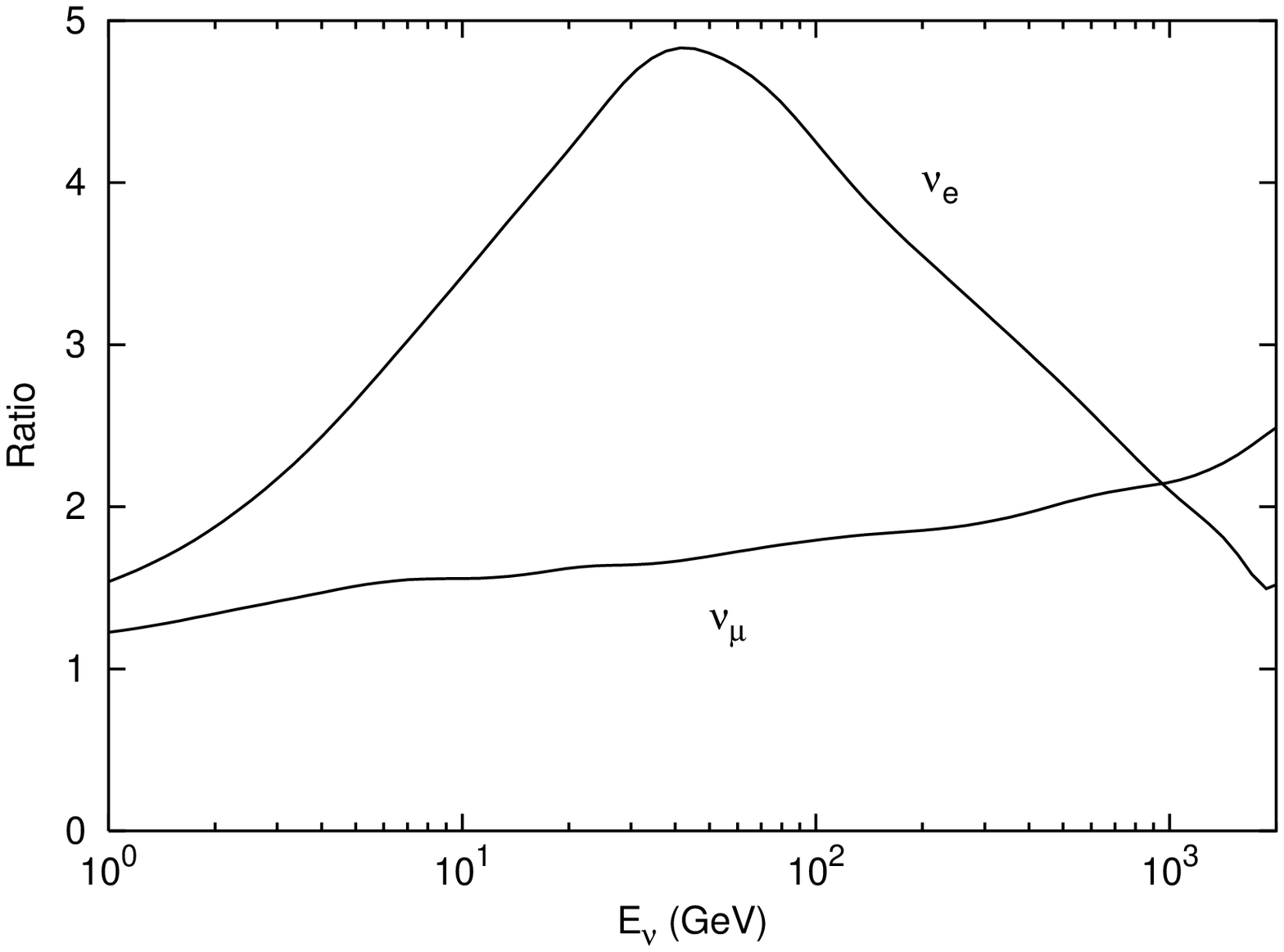,width=8.cm}}
\label{fig9}
\vspace{-1cm}
\caption{Ratio of horizontal ($0.<|\cos\theta|<0.375$)
to vertical ($0.675<|\cos\theta|<1.$) atmospheric neutrinos.
}
\end{figure}

\subsection{Flavor and charge ratios}

The primary source of electron neutrinos is muon decay, which also
gives an equal contribution of muon neutrinos.  For 
$E_\mu>2~{\rm GeV}/\cos\theta$ muons on average reach the ground
before they decay.  For $\theta > 70^\circ$ the curvature of the
Earth limits the pathlength in the atmosphere to a few hundred km and
muons with $E_\mu> 30$~GeV typically reach the ground before
decaying.   For $E_\nu$ more than $\sim100$~GeV/$\cos\theta$,
the contribution from muon decay practically vanishes and
the ratio of $\nu_e/\nu_\mu$ approaches a small value $\lesssim 5$\%,
with most of the $\nu_e$ coming from $K_{e3}$ decays.  Somewhere
above a TeV, charm decay dominates $\nu_e$.

Atmospheric neutrinos constitute the primary background, as well
as a useful calibration source, for neutrino telescopes.
The characteristic angular and energy dependences of the fluxes
of $\nu_e$ and $\nu_\mu$ described here
should be useful for calibration.  As an example, Fig.~9 
shows the ratio of horizontal to vertical neutrinos for
two flavors from Ref.~\cite{AGLS}  The angular asymmetry
of muon neutrinos increases steadily with energy to a ratio $\sim2$
in the TeV range.  The large horizontal to vertical ratio 
for $\nu_e$ and its energy-dependence above $100$~GeV should be
a distinctive calibration signature in detectors large enough
to measure the low rate of $\nu_e$-induced cascades.

The ratio $\nu_\mu/\bar{\nu}_\mu$ is also of interest.
As magnetized detectors such as MINOS~\cite{minos} begin
to operate, it will be possible to measure the charge
ratio of atmospheric neutrino-induced muons.
At the high energies
the kaon contribution will be of great importance for 
evaluating the $\nu/\bar{\nu}$ ratio, which, together with
the cross sections $\sigma_\nu > \sigma_{\bar{\nu}}$, determines
the muon charge ratio.  Because the contribution of $\nu_\mu$
is larger than that of $\bar{\nu}_\mu$,  $\mu^+/\mu^- < 1$.

\noindent
\begin{figure}[t]
%\vspace*{-1cm}
\flushleft{\epsfig{figure=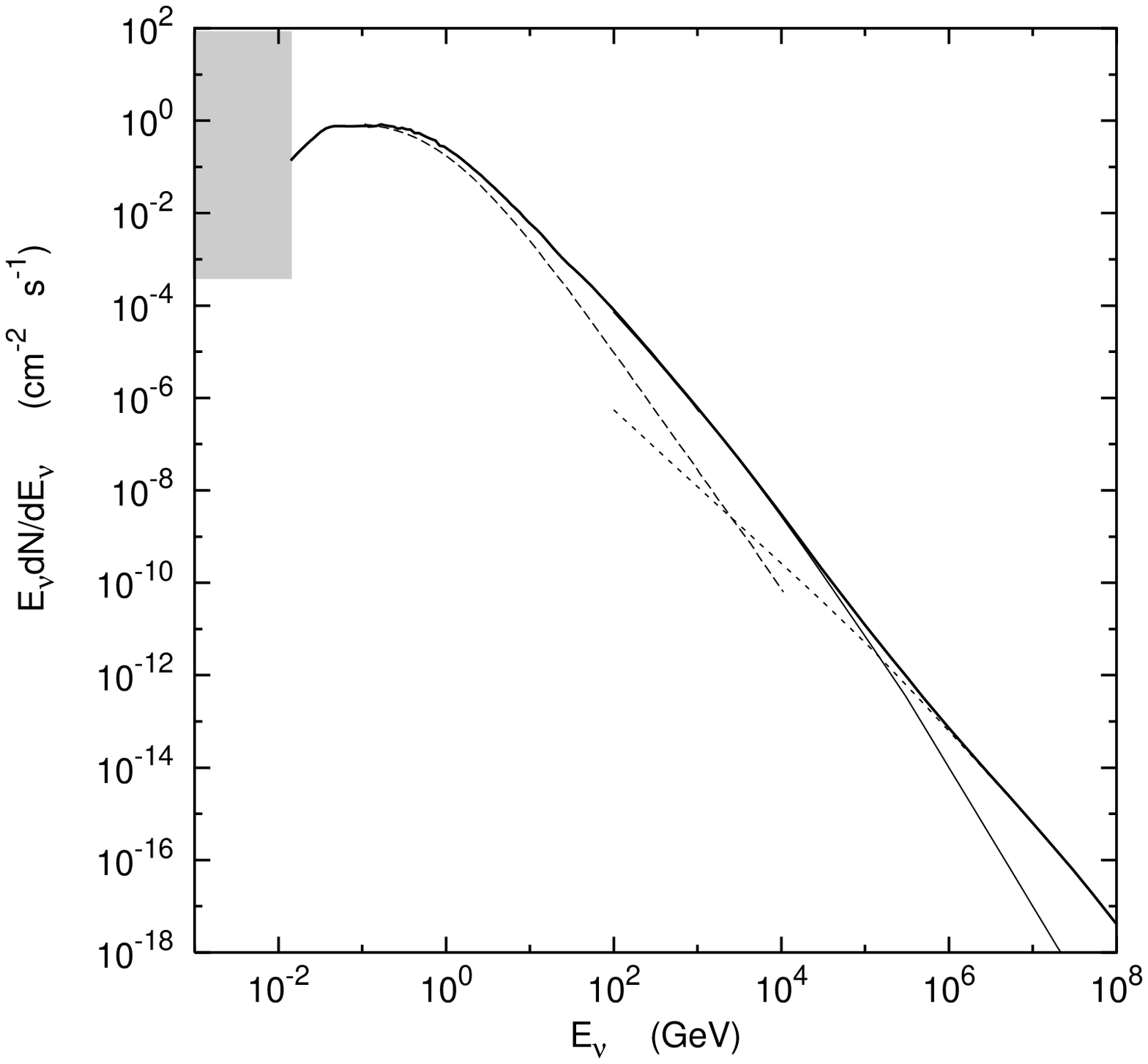,width=8.cm}}
\label{fig10}
\vspace{-1cm}
\caption{Global view of the atmospheric neutrino spectrum integrated
over all directions.  The upper solid line shows the total flux of 
$\nu_\mu+\bar{\nu}_\mu$ including the effects of oscillations.
Above 100 TeV this flux is dominated by prompt neutrinos.
Broken lines are $\nu_e+\bar{\nu}_e$, with the crossover to 
prompt around 1 TeV.  Shaded area indicates solar neutrinos.}
\end{figure}

\subsection{Prompt neutrinos from charm}

Leptonic decays of short-lived charmed hadrons gives rise
to a third term (not shown) on the right side of Eq.~\ref{analytic},
of the same form as the term for the kaon contribution.
The spectrum-weighted moment $Z_{N\rightarrow {\rm charm}}$ is
much smaller, but $\epsilon_{\rm charm} \gtrsim 10^7$~GeV, so this
component dominates the neutrino spectrum at sufficiently high energy.
There are significant uncertainties in inclusive cross sections
for hadronic production of charm and hence for the prompt
neutrino flux.~\cite{Costa}  Because
of the steep cosmic-ray energy spectrum, charm production
in the forward fragmentation region is likely to make an
important contribution to the flux of prompt leptons,
in addition to production via central QCD processes.
Fig.~10 shows a global view of the spectrum
with an estimate of charm production from Ref.~\cite{Bugaev}.
In this quark-parton recombination model, 
charm becomes the dominant source of $\nu_e$ for $E_\nu\gtrsim 2$~TeV
and of $\nu_\mu$ for $E_\nu\gtrsim 100$~TeV.  Prompt neutrinos
have an isotropic angular distribution.  The increasing importance
of prompt $\nu_e$ contributes to the rapid decrease
of the ratio of horizontal to vertical electron neutrinos
above $100$~GeV in Fig.~9.

\noindent
{\bf Acknowledgments} I thank M. Honda for collaboration on this
work, Todor Stanev for reading the manuscript and Doug Michael and
Peter Litchfield for helpful discussions.

\end{document}